\documentclass[preprint,12pt]{aastex}

\newcommand{\degree}{\mbox{$^{\circ}$}}
\newcommand{\am}{\mbox{\arcmin}}
\newcommand{\as}{\mbox{\arcsec}}

\newcommand\cmv{\mbox{cm$^{-3}$}}

\def\lsim {$\rlap{\raise.4ex\hbox{$<$}}\lower.55ex\hbox{$\sim$}\,$}

\newcommand{\lsun}{\mbox{L$_\odot$}}

\newcommand{\mean}[1]{\mbox{$\langle#1\rangle$}} 
\newcommand{\av}{\mbox{$A_V$}} 
 
\input{epsf}

\newcommand{\kaprat}{\mbox{$\kappa_{smm}/\kappa_{ir}$}}
\newcommand{\kapratK}{\mbox{$\kappa_{smm}/\kappa_{2.2}$}}

\begin{document}

               
\title {\bf Observational Constraints on Submillimeter Dust Opacity}
\author{Yancy L. Shirley\altaffilmark{1,2},
        Tracy L. Huard\altaffilmark{3},
	Klaus M. Pontoppidan\altaffilmark{4,8},
        David J. Wilner\altaffilmark{5},
	Amelia M. Stutz\altaffilmark{6},
        John H. Bieging\altaffilmark{1},
        and Neal J. Evans II\altaffilmark{7},
	}
\altaffiltext{1}{Steward Observatory, 933 N Cherry Ave., Tucson, AZ 85721}
\altaffiltext{2}{Adjunct Astronomer, The National Radio Astronomy Observatory}
\altaffiltext{3}{Astronomy Department, University of Maryland, College Park, 
MD 20742}
\altaffiltext{4}{Caltech, Pasadena, CA}
\altaffiltext{5}{Harvard-Smithsonian Center for Astrophysics, 60 Garden Street, 
Cambridge, MA 02138}
\altaffiltext{6} {Max-Planck-Institut f\"ur Astronomie, K\"onigstuhl 17,
D-69117 Heidelberg, Germany}
\altaffiltext{7}{University of Texas at Austin, 1 University Station C1400, 
Austin, TX 78712-0259}
\altaffiltext{8}{Hubble Fellow}
 
\begin{abstract}

Infrared extinction maps and submillimeter dust continuum maps
are powerful probes of the density structure in the envelope of star-forming 
cores.  We make a direct comparison between infrared and submillimeter dust 
continuum observations of the low-mass Class 0 core, B335, to 
constrain the ratio of submillimeter to infrared opacity (\kaprat ) and
the submillimeter opacity power-law index ($\kappa \propto \lambda^{-\beta}$).
Using the average value of theoretical dust opacity models
at $2.2$ \micron , we constrain the dust opacity at $850$
and $450$ \micron .  
Using new dust continuum models based upon the broken
power-law density structure derived from interferometric observations of B335
and the infall model derived from molecular line observations
of B335, we find that the opacity ratios are
$\frac{\kappa_{850}}{\kappa_{2.2}} =  (3.21 - 4.80)^{+0.44}_{-0.30} \times 10^{-4}$
and
$\frac{\kappa_{450}}{\kappa_{2.2}} =  (12.8 - 24.8)^{+2.4}_{-1.3} \times 10^{-4}$
with a submillimeter opacity power-law index of
$\beta_{smm} = (2.18 - 2.58)^{+0.30}_{-0.30}$. 
The range of quoted values are determined from the uncertainty in the physical
model for B335. 
For an average $2.2$ \micron\ opacity of $3800 \pm 700$ cm$^2$g$^{-1}$,
we find a dust opacity at 850 and 450 \micron\ of 
$\kappa_{850} = (1.18 - 1.77)^{+0.36}_{-0.24}$ and 
$\kappa_{450} = (4.72 - 9.13)^{+1.9}_{-0.98}$ cm$^2$g$^{-1}$ of dust.
These opacities are from $(0.65 - 0.97) \kappa^{\rm{OH}5}_{850}$ of
the widely used theoretical opacities of Ossenkopf and Henning for
coagulated ice grains with thin mantles at $850$\micron .  

\end{abstract}

\keywords{stars: formation  --- ISM: dust, extinction --- ISM: clouds ---}


\section{Introduction}

The commissioning over a decade ago of two dimensional 
bolometer cameras such as 
SCUBA (Submillimeter Common User Bolometer Array; Holland et al. 1999)
and SHARC (Submillimeter High Angular Resolution Camera; Hunter et al. 1996)
permitted the efficient mapping of
dust continuum emission at submillimeter wavelengths.  
Submillimeter continuum observations of star forming regions in the Milky Way
have constrained the physical structure of dense, star-forming
cores through all embedded phases of core and protostar
formation (e.g., Shirley et al. 2002; J\o rgensen et al. 2002; Williams,
Fuller, \& Sridharan 2005, Kirk et al. 2005).  On larger scales, mapping of
entire molecular clouds has constrained the
the dense core Initial Mass Function (IMF) and made surprising
connections to the shape of the stellar IMF (e.g., Motte et al. 1998,
Johnstone et al. 2001, Enoch et al. 2007, Motte et al. 2007).  
In all of these studies, the
mass or density scale is set by the assumed submillimeter dust 
opacity $\kappa$ (cm$^2$g$^{-1}$) since  
the mass of optically thin submillimeter 
emission is inversely proportional to the dust opacity 
($M_d \propto 1/\kappa$; Hildebrand 1983).
It is very important to use an accurate value of the dust
opacity since the mass distribution within the protostellar core
directly affects the dynamical stability of the core as well as the 
radiative transfer through the core.
The uncertainty in submillimeter dust opacity is the largest source of 
uncertainty in mass calculations and radiative transfer models of 
protostellar cores (see Shirley et al. 2005).  

Since the pioneering work of Knacke \& Thompson (1973), there have
been many attempts to estimate the submillimeter dust
opacity in star forming cores. 
The opacity of dust grains in the general ISM
can vary substantially from dust opacities in environments surrounding
star formation.  From optical studies of dust absorption,
it is well known that the ratio of total to selective extinction,
$R_V = A_V / E(B-V)$, 
varies from $R_V = 3.1$ in the general ISM to larger
values ($R_V = 5.5$) in denser star forming regions
(e.g., Mathis 1990, Whittet 2003, Draine 2003).  Several important
physical  processes directly affect the opacities
in dense regions (Henning, Michel, \& Stognienko 1995).
The compositions of dust grains may vary from region to region.
Dust grains may coagulate, changing the shape and the 
normalization of the general ISM size distribution (e.g., 
Mathis, Rumpl, \& Nordsieck 1977; Mathis, Mezger, \& Panagia 1983,
Ossenkopf 1993, Ormel et al. 2009).  
Most of the molecular gas in a dense core is shielded 
from energetic photons from
the forming protostar or from the Interstellar
Radiation Field (ISRF) and confined to temperatures
less than $20$K.  Many molecular species freeze out
of the gas phase forming layers of polar (H$_2$O) and 
apolar (CO) ices that change the dielectric properties of
the grains and the size distribution of grains.  
Theoretical calculations of the opacities
make various assumptions about grain composition,
grain size distributions, grain geometry, grain porosity,
and ice mantle compositions; the resulting predicted submillimeter 
opacities vary by up to an order of magnitude (see Table 2 of
Shirley et al. 2005).  Currently, the most widely used
calculation of opacities is that of Ossenkopf \& Henning (1994)
which takes into account coagulation and varying thicknesses
of ice mantles for dust grains that have persisted 
at high density, $n = 10^6$ \cmv , for $10^5$ years.  
Even the Ossenkopf \& Henning opacities vary by a factor
of a few at submillimeter wavelengths depending on the
particular assumptions used in the model and there is no guarantee
that these opacities are appropriate for the wide variety of
environments towards which they have been applied, 
from cold, low-mass starless cores to warm, high-mass
proto-cluster cores.

Clearly, observational constraints on submillimeter dust opacities are 
needed. There have been a few attempts to constrain the
dust opacity at long wavelengths.  These methods use
observations of the amount of dust extinction at near-infrared
wavelengths directly compared to the amount of dust emission
at (sub)millimeter wavelengths
to constrain the opacity ratio between submillimeter
and near-infrared (or visible) wavelengths.  Kramer et al.
(1998,2003) studied the opacity ratio in several dense
starless cores in the IC 5146 region.  Similarly,
Bianchi et al. (2002) studied the nearby low-mass
starless core B68.  Both groups have demonstrated that
the technique works if sensitive observations are obtained
at both submillimeter and near-IR wavelengths and care is taken to 
compare observations 
taken at two very different resolutions (pencil beam at near-infrared
wavelengths versus beam convolved emission at submillimeter wavelengths
with non-Gaussian telescope power patterns).  This method has never been 
applied before to a dense core which harbors a protostar 

The near-infrared to submillimeter comparison requires high
signal-to-noise observations at three wavelengths (two
near-infrared wavelengths and one submillimeter wavelength).
Few submillimeter studies of protostellar regions have
sensitive maps that detect extended emission at high
signal-to-noise ratios over large regions of the core ($> 2$\arcmin )
because the ground-based observations are limited by
the size of the chopping needed for sky subtraction.
One promising object is the nearby, well studied Class 0
protostar located within the dense isolated Bok globule, Barnard 335. 
Since its initial detection as a far-infrared source
(Keene et al. 1983), B335 has received considerable
attention as it is one of the best protostellar infall candidates
(Zhou et al. 1993, Choi et al. 1995, Evans et al. 2005) identified
during the deeply embedded phase of low-mass star formation.  
For the purposes of this study, B335 has been observed with high
sensitivity with SCUBA at
submillimeter wavelengths (Shirley et al. 2000) and with
NICMOS at near-infrared wavelengths (Harvey et al. 2001).

In this paper, we utilize a method of comparing the near-infrared
extinction to submillimeter emission to constrain the dust
opacity ratio between submillimeter wavelengths and $2.2$ \micron\
(\S2) toward B335.  From dust continuum radiative transfer 
we derive an updated physical model for B335 using currently
published dust opacities (\S3).  The dust opacity ratio is
determined utilizing our constraints on $T_d(r)$ (\S4.1).  We then constrain
the opacity at submillimeter wavelengths and the opacity
power law index, $\beta_{smm}$ (\S4.2,4.3).  Throughout this paper,
we refer to the dust mass opacity, $\kappa_{\nu}$, in units
of cm$^2$ per gram of dust.

\section{Method}

By directly comparing infrared extinction to submillimeter emission maps
along common lines of sight, 
we constrain the opacity ratio
\kaprat .  
The emission at submillimeter wavelengths along a line-of-sight, $s$, is given by the
equation of radiative transfer for optically thin emission, 
$dI_{\nu}/d\tau_{\nu} = B_{\nu}[T(s)]$ with 
\begin{equation}
\frac{d\tau_{smm}}{ds} = \mu \, m_{\rm{H}}\, 
\langle m_d/m_g \rangle \kappa_{smm}(s)\, n(s) \;\; ,
\end{equation}
where $B_{\nu}$ is the Planck function and $n(s)$ is the gas number density (\cmv ).
We use the dust mass opacity, $\kappa_{\nu}$ (cm$^2$g$^{-1}$), 
where we have assumed a mean molecular
weight $\mu = 2.32$ and an average gas mass to dust mass ratio of 
$\langle m_d/m_g \rangle = 1:100$.  In the derivation of the expression
to determine the
opacity ratio, $\mu$ and $\langle m_d/m_g \rangle$
cancel out; but they are stated explicitly here since they are 
used in radiative transfer modeling of the submillimeter emission (\S3).
Integrating along the line-of-sight gives
\begin{equation}
I_{smm} = \int_s B_{\nu}[T(s)] \mu m_{\rm{H}} \langle m_d/m_g
\rangle n(s) \kappa_{smm}(s) \, ds \;\; ,
\end{equation}
(Adams 1991, Shirley et al. 2000).

At infrared wavelengths, the observed intensity 
is due to the total amount of extinction along the
line-of-sight, 
\begin{equation}
I_{ir}(s) = I_{ir}(0) e^{-\tau_{ir}} \;\; ;
\end{equation}
therefore, the total extinction in magnitudes is
\begin{equation}
A_{ir} = -2.5 \log \left( \frac{I_{ir}(s)}{I_{ir}(0)} \right) \; = 2.5 \log(e) \int_s \mu 
m_{\rm{H}} \langle m_d/m_g \rangle 
n(s) \kappa_{ir}(s) ds  \;\; .
\end{equation}
The infrared opacity includes contributions from absorption and scattering
($\kappa_{ir} = \kappa_{ir}^{abs} + \kappa_{ir}^{sca}$).
Dividing Equation (2) by Equation (4), we derive the relationship
between the submillimeter intensity, the infrared extinction, and the opacity ratio
\begin{equation}
I_{smm} = \frac{\int_s B_{\nu}[T(s)] n(s) 
\kappa_{smm}(s) ds}
{2.5 \log(e) \int_s n(s) \kappa_{ir}(s) ds } \; A_{ir} \;\; .
\end{equation}
If we further assume that the infrared and submillimeter opacity does not vary
along the line-of-sight, then we find 
\begin{equation}
\frac{I_{smm}}{P_n} =  \left( \frac{\kappa_{smm} }{\kappa_{ir} } \right) \; A_{ir} \;\; ,
\end{equation}
where $P_n$ is related to the density-weighted average Planck function and
is defined as
\begin{equation}
P_n = \frac{\int_{s} n(s) B_{\nu}[T(s)] \, ds}{2.5 \log(e) \int_{s} n(s) \, ds} \;\; .
\end{equation}
If an isothermal approximation is used, then the column density
cancels in $P_n$ and we are left with $P_n = B_{\nu}(T_{iso})/2.5 \log(e)$.
However, dust temperature gradients exist throughout protostellar envelopes
(Shirley et al. 2003).  In this paper, we shall use the $n(r)$ and $T_d(r)$
determined from the best-fitted 1D dust continuum radiative transfer models
to calculate $P_n$ along each stellar line-of-sight.

Theoretically, the opacity ratio is determined from the slope of
a plot of the submillimeter intensity versus the near-infrared
extinction.  In reality, neither the submillimeter intensity
nor the near-infrared extinction are directly observed.
Submillimeter observations actually observe the convolution of
the source specific intensity distribution with the telescope
beam pattern such that the observed flux density is 
$S_{smm} = \mean{I_{smm}} \Omega_{beam}$.  
A typical submillimeter beam pattern is
not well described with a single Gaussian main beam as
a significant fraction of the power pattern is contained within the 
sidelobes.  The total solid angle of the beam, $\Omega_{beam}$,
is determined from the integral of the normalized telescope
power pattern, including sidelobes, over solid angle.

With observations of 
two or more near-infrared wavelengths, we observe color differences of 
background stars.  The extinction along a line of sight is determined 
from a scaling law that relates $A_{ir}$ to the observed
color excess.  In this paper, we will determine the 
extinction at the near-infrared K band at $2.2$ \micron .
We define the K band selective extinction, $R_K$ such that
\begin{equation}
A_{2.2} = R_K [ (H - K) - \mean{(H - K)_0} ] = R_K E(H - K) = R_K (A_{1.65} - A_{2.2}) \, ,
\end{equation}
where $\mean{(H - K)_0}$ is the mean intrinsic (H-K) color of
the background stars and $(H - K)$ is the observed
infrared colors with extinction.
The determination of $R_K$ 
assumes that the near-infrared extinction law is well described 
by a power-law ($A_{ir} \propto \lambda^{- \beta_{nir}}$ for near-infrared
wavelengths; see
Draine 2003 and references therein, also Flaherty et al. 2007).  
Chapman \& Mundy (2009) determine $\beta_{nir} = 1.7$ from the slope of a plot
of $E(J - H)$ versus $E(H - K)$ of background stars toward four dense cores.
This is in the middle of the range of the 
typical values of $\beta_{nir} = 1.6 - 1.8$ found in the literature
(Draine 1989, Rieke \& Lebofsky 1985, Martin \& Whittet 1990, 
Whittet et al. 1993).  Using this range of $\beta_{nir}$ in Equation (8),
we find $R_K = 1.59 \pm 0.12$.  We shall use this value of $R_K$
throughout this paper.

We constrain \kapratK\ from the slope of a plot of $S_{smm}/P_n
\Omega_{beam}$ versus $E(H - K)$.  If the slope is denoted by $b$, then
the opacity ratio is simply \kapratK\ $= b/R_K$. 
It is important to reiterate that this derivation assumes that the 
dust opacity 
ratio along an individual line-of-sight is constant. If there is
a variation in the dust opacity, then  a linear regression is
measuring an emission-weighted average dust opacity ratio along each line-of-sight.
A monotonic change in the opacity ratio with radius will produce curvature 
in a plot of $S_{smm}/P_n \Omega_{beam}$ versus $E(H - K)$
while a distribution of opacity ratios along each line-of-sight 
will produce intrinsic scatter in the correlation. 


\subsection{Submillimeter Images}

The reduction and analysis of SCUBA 850 and 450 \micron\
jiggle maps is described in detail by Shirley et al. (2000).
We have re-analyzed and combined SCUBA images of B335 taken by
Shirley et al. (2000) and images from the SCUBA CADC archive.
B335 was observed with SCUBA in jiggle mapping mode on only two nights 
(April 17, 1997 and December 18, 1997) with low atmospheric opacity 
($\tau_{225} < 0.05$).  Throughout this paper, for simplicity
we quote the SCUBA wavelengths as 850 and 450 \micron , whereas the 
actual narrow-band SCUBA filters have average wavelengths 
of 860 and 445 \micron\ respectively.  The average
wavelengths are not sensitive to the shape of the source spectrum
(i.e. the average wavelength changes from 860 to 859 \micron\ 
for sources with $\nu^0$ to $\nu^4$.)

Before combining images from two different nights, observations of Uranus 
taken within 1 hour before and after the B335 observations on each night
were analyzed to compare the shape of the telescope beam
pattern.  
We found no significant difference in the Uranus radial profiles between
the two sessions.   The flux density scale is
calibrated using the peak and integrated Uranus flux 
observed on the same night 
as the core as determined from the FLUXES program 
\footnote{http://www.jach.hawaii.edu/jac-bin/planetflux.pl}  
We found very good agreement in the peak flux on both nights.
The specific intensities in the final image (erg s$^{-1}$ cm$^{-2}$
Hz$^{-1}$ sr$^{-1}$) were estimated from the measured flux densities
by using the average of the solid angle of telescope beam pattern
determined on the same nights as the B335 observations 
($\langle\Omega_{850}\rangle = 7.3 \times 10^{-9}$ sr and 
$\langle\Omega_{450}\rangle = 4.4 \times 10^{-9}$ sr).  
The rms noise in the combined images
is $15$ mJy/beam at $850$ \micron\ and $82$ mJy/beam at $450$ \micron .
The combined images are very similar to the published contour maps
in Shirley et al. (2000; see Figure 1).

The final image pixels are over-sampled ($1$\as ) to determine
the closest lines-of-sight for comparison with background stars (Figure 1).
Since multiple background stars may lie within a single SCUBA beam
($15$\as\ at 850 \micron\ and $8$\as\ at 450 \micron ),
the submillimeter intensity will be semi-correlated in a plot
of $S_{smm}/P_n\Omega$ versus $E(H - K)$.  By using the
oversampled SCUBA map, we preserve all of the information in
the near-infrared extinction map.  Smoothing the near-infrared line-of-sight
to a $20$\as\ $-$ $30$\as\ resolution, as is typically done with
extinction mapping methods (e.g. Teixeira et al. 2005; Alves et al. 2001),
results in a suppression of the true scatter in line-of-sight
$E(H - K)$.  For this reason, we do not smooth the near-infrared data.
Similarly, attempting to deconvolve the SCUBA beam from the
B335 image is extremely difficult because the two
dimensional beam shape varies during the observations due to changes
in the shape of the telescope surface and variation of parallactic angle.

\subsection{Near-infrared Images}

The near-infrared images are from NICMOS observations and are
analyzed in detail by Harvey et al. (2001).  Observations were
performed with the NIC3 camera on the \textit{Hubble Space Telescope} 
using the F160W and F222M filters.  A detailed photometric comparison
was made between the NICMOS magnitudes and J, H, and K
observations made with the NIRC camera on Keck I.    
B335 was imaged with a
$3 \times 3$ mosaic (NIC3 FOV 51\farcs 2)
plus a $4$\am\ radial strip centered on the Class 0 protostar.
Background fields, off the Barnard 335 cloud, at similar
galactic latitude were observed with NICMOS and NIRC.
The mean intrinsic color of background 
stars is $\mean{(H - K)_0} = 0.13$ mag and
intrinsic scatter in the color is $\sigma(H -  K)_0 = 0.16$ mag.

The distribution of background stars compared to $850$ \micron\
emission is shown in Figure 1.  More than $200$ background
stars are observed.  Most of the background stars are located outside
of $20$\as\ with only 4 stars observed within $20$\as\
of the protostar and none within $15$\as .  The inability to see
background stars in the innermost regions of Barnard 335 is
due to the high column densities observed toward the core.  Therefore,
a submillimeter to near-infrared opacity comparison can
only be made in the outer regions of the envelope greater than
$15$\as\ from the protostar ($3750$ AU projected 
at a distance of $250$ pc).

\section{An Updated 1D Dust Model for B335}

The density and dust temperature along each line-of-sight are determined from
radiative transfer models of the dust continuum emission (e.g., 
Shirley et al. 2003).
The radiative transfer models self-consistently calculate the dust temperature 
profile, $T_d(r)$, using a one dimensional radiative transfer
code (CSDUST3; Egan, Leung, \& Spagna 1988) determined from an input density distribution
($n(r)$), interstellar radiation field, internal luminosity, and a 
dust opacity curve ($\kappa(\lambda )$).  
The model intensity profiles and spectral energy distribution are reconstructed
using the same techniques as the observations (e.g., beam convolution, chopping,
aperture matching).  The best-fitted models minimize the $\chi^2_r$  for 
the observed submillimeter 
intensity profiles and the observed spectral energy distribution at
wavelengths where the optical depth is less than unity
($\lambda > 60$ \micron ).  Details of the radiative
transfer modeling procedures for low-mass cores 
may be found in Shirley et al. (2002, 2005).

Unfortunately, in order to calculate $P_n$ and ultimately the observed
opacity ratio, we have to assume a dust opacity curve to input into
the radiative transfer model.  Therefore, we must explore different
theoretical opacities to determine how our derived opacity ratio is biased by our 
radiative transfer opacity choice.  Since the background stars toward B335
are at projected lines-of-sight greater than $1000$ AU from the protostar
and since B335 is a low luminosity protostar ($L_{bol} = 3.3$ \lsun ) embedded
in a dense core that is exposed to a weaker than average ISRF (Shirley et al. 2002; Evans
et al. 2005), the
calculated $T(r)$ at those radii will be a slowly varying quantity with radius
(Shirley et al. 2002).  In this paper, 
we explore the effect on $T_d(r)$ for different 
dust opacity models calculated by 
Ossenkopf \& Henning for coagulated grains with 
varying thicknesses of ice mantles 
(columns 2,5, and 8 of Ossenkopf \& Henning 1994) and dust opacity models
calculated by Weingartner \& Draine (2001) for ISM grain populations
with $R_V =$ 3.1, 4.0, and 5.5 (size distribution ``A''\footnote{see http://www.astro.princeton.edu/$\sim$draine/dust/dustmix.html for the
latest versions of these opacities.}).  
Since the original Ossenkopf \& Henning models
did not calculate the scattering opacity, the OH opacity was divided
between scattering and absorption using the ratios from
the Pollack et al. (1994) models that
best match the Ossenkopf \& Henning absorption opacity (Young \& Evans
2005).  The ratio of scattering to absorption opacities across the 3 $\mu$m
ice feature were determined from the albedos 
in Pendelton et al. (1990, Figure 4b) since the Pollack et al.
models did not include ices. 
A complete explanation of the modifications to the short wavelength OH 
opacities may be found in \S2.1 of Young \& Evans (2005).
These short wavelength-modified OH opacities have been used
in several published dust continuum radiative transfer calculations
(e.g., Young \& Evans 2005, Shirley et al. 2005, Dunham et al. 2006,
Dunham et al. 2010).

We assume a distance of $250$ pc
(Tomita et al. 1979) to be consistent with previously published models, 
although this distance is very uncertain (see Olofsson \& Olofsson 2009).  
Shirley et al. (2002) explored the effects of a closer distance (125 pc) on the radiative 
transfer models; however, the determination of the opacity ratio 
does not depend on distance.

There is also uncertainty in the modeled density structure of the B335 core.
The original published dust continuum models of B335 found that a 
single power-law density model ($n(r) \propto r^{-1.8}$)
using OH5 dust opacities
is a good fit to the observed intensity profile while the fit to the shape 
of the far-infrared SED was not as well matched (Shirley et al. 2002).
Subsequently, Harvey et al. (2003a,b) published an updated
model based on interferometric millimeter continuum emission and near-infrared
extinction maps (Harvey et al. 2001) and found that a broken
power-law was a better fit to the density structure with a 
flatter density profile ($r^{-1.5}$) inside 6500 AU and steeper
profile outside ($r^{-2.0}$).  Most recently, Doty et al. (2010) 
used broken power-laws and variable opacities in four radial
zones to attempt to model the large-scale dust continuum emission
(excluding interferometric constraints).
All of these modeling efforts found strong disagreement with the previously
published molecular line radiative transfer modeling which 
is best-fit by a Shu (1977) infall solution with a modest infall
radius $r_{inf} = 6200$ AU (Choi et al. 1995).  Evans et al. (2005)
confirmed this disagreement with updated radiative transfer modeling
using non-uniform abundance profiles for several molecular
species.  Reconciling the modeling differences requires simultaneous
modeling of molecular line and dust emission (interferometric and
single-dish) with varying dust opacities with radius 
(e.g., including CO desorption for $T_d > 20$ K).  Unfortunately, this
is beyond the scope of this paper; therefore, we
shall analyze the uncertainty on the opacity ratio due to
our uncertainty in the underlying density model by using both
the best-fit molecular line model and the best-fit dust continuum model.

We explore a grid of models that varies the dust opacity
and the density scale factor, $f$.  The entire density profile is scaled
by a single number, $f$, to match the observed flux in a $120$\as\ aperture 
at $850$ \micron .   The strength of the ISRF is constrained from molecular line modeling 
of CO observations (Evans et al. 2005).  
We adopt an ISRF parametrized in Shirley et al. (2005) and 
corresponding to $G_0 = 0.1$ Habings
($s_{isrf} = 0.3$, \av $(R_{o}) = 1.0$ mag)
at an outer radius of $R_o = 3 \times 10^4$ AU, consistent 
with the extent of the near-infrared extinction profile (Harvey et al. 2001).
Two physical models are used: a scaled Shu (1977) infall solution
($n(r) = f n_{\rm{Shu}}(r)$) with $r_{inf} = 6200$ AU; and a broken
power-law solution derived by Harvey et al. (2003a,b),
\begin{eqnarray}
n(r) & = & 3.3 \times 10^{4} f \rm{cm}^{-3} \left( \frac{r}{6500 \rm{AU}} \right)^{-1.5} \;\; r \in [100,6500] \, \rm{AU} \\
n(r) & = & 3.3 \times 10^{4} f \rm{cm}^{-3} \left( \frac{r}{6500 \rm{AU}} \right)^{-2.0} \;\; r \in [6500,30000] \, \rm{AU} \;\;.
\end{eqnarray}
Results for a subset of the models are summarized in Table 1.

As noted by Shirley et al. (2002), the scaled Shu infall models do not
fit the observed submillimeter intensity profiles or the SED.  
In contrast, the broken power-law of Harvey et al.
scaled in density by a factor of $f = 2.4$ with OH8 opacities provide
a good fit to the submillimeter intensity profiles and bolometric
luminosity and a slightly better fit to the shape of the SED than the originally published
best-fitted model by Shirley et al. (2002; Figure 2) which use OH5
opacities.  All of the radiative
transfer models have difficulty fitting the far-infrared SED 
indicating that there may be a problem with the theoretical opacities 
at those wavelengths or effects of non-spherical geometry 
for B335.  Since the Weingartner \& Draine opacities
are much smaller at submillimeter wavelengths than the Ossenkopf \& Henning
opacities, the density scaling factor is higher for WD models.
We calculate $P_n(\theta)$ for all of the models listed in Table 1 to analyze 
the effect of the model opacities on the derived opacity ratio (\S4.2).

\section{Results}

\subsection{Linear Regression Technique}

     In order to determine the opacity ratio between a submillimeter
wavelength and $2.2$ \micron , we must determine
the slope from the plot of submillimeter intensity versus 
the near-infrared color excess.  We use the Bayesian linear
regression routine LINMIX\_ERR (Kelly 2007) to determine the
slope of a relationship of the form
\begin{equation}
y = a + bx + \sigma_{int}^2
\end{equation}
with intrinsic scatter about the line, $\sigma_{int}^2$, and heteroscedastic
errors in both x and y.  In our analysis, $x = E(H - K)$
and $y = S_{smm}/P_n \Omega_{beam}$.
LINMIX\_ERR approximates
the distribution of the independent variable ($x$) as a mixture of Gaussians.
This method alleviates the ad hoc assumption of a uniform prior distribution
on the independent variable
that is used in the derivation of popular $\chi^2$ minimization routines 
such as XYEFIT (e.g., Press et al. 1992, Tremaine et al. 2002, 
Weiner et al. 2006) and
also permits fitting of truncated data sets 
(e.g., Malmquist bias) and data sets
that include censored data or upper limits.
Details of the assumed prior distributions are described in detail in 
Kelly (2007).
Direct computation of the posterior distribution is too computationally 
intensive; therefore, random draws from the posterior distribution
are obtained using a Markov Chain Monte Carlo method 
(Metropolis-Hastings Algorithm, Chib \& Greenberg 1995; Tierney 1998;
Gelman et al. 2004).  
We fit our
data using the publicly available IDL code LINMIX\_ERR.pro (Kelly 2007) 
to determine
the distributions of $a$, $b$, and $\sigma_{int}^2$.  In order to test the robustness of
the resulting distributions, we varied the number of Gaussians
from k=2 to k=4 and used various numbers of iterations up to $10^4$.

The plots of $S_{smm}/P_n \Omega_{beam}$ versus $E(H-K)$ are shown
in Figures 3 and 4 for the best-fitted dust continuum model ($2.4\times$Harvey
BPL OH8) at $850$\micron\ and $450$\micron .  
The histograms of the slope and intercept are well approximated
by Gaussian distributions (see inset of Figures 3 and 4); therefore,
we tabulate the mean and standard deviations of the 
slope and intercept distributions (Table 1).
The linear regression is only performed on data points that
lie outside the outflow cavities.  Harvey et al. (2001) noticed a 
bimodal distribution of color excesses depending on whether a
background star was located in the direction of the east-west
oriented molecular outflow cavity.  The opening angle of the CO
outflow was recently characterized by Stutz et al. (2008) to
be $55$\degree .  This is slightly larger than the outflow opening
angle of $40 \pm 5$\degree\ used in Harvey et al. (2001) and (2003a).
Since we are making a comparison of the opacity properties
of the dust in the envelope, and since shock processing of dust within the outflow
cavity walls may affect the grain opacities, we use the larger
estimate for the outflow opening angle ($55$\degree ) in this paper.

\subsection{Opacity Ratios and Submillimeter Opacity}

	The opacity ratio is determined from plots similar to
Figures 3 and 4 for all of the models discussed in \S3.  Linear
regression from the LINMIX\_ERR method provide good fits to
the observed correlations.  We find no evidence for large, systematic
curvature in the plots of $S_{smm}/P_n\Omega$ versus
$E(H - K)$ indicating that over the range of impact parameters
probed by background stars ($\theta \in [15$\as $,70$\as $]$),
there is not a strong monotonic gradient in the intensity-weighted
dust opacity with radius.  This result independently confirms the conclusions
by Shirley et al. (2002) that there is no evidence for large
scale opacity changes from comparisons of $850$ and $450$ \micron\
model intensity profiles.  This result is also consistent
with the findings of Doty et al. (2010) which indicate nearly
constant dust opacities over the range of radii that are fit in
this paper.  A caveat is that there is significant scatter in
the correlations at both submillimeter wavelengths that could
mask opacity changes.  The scatter
appears to become slightly larger near the highest $E(H - K)$
observed.  This may indicate that opacities are beginning
to change within a few thousand AU of the protostar.  Theoretically,
we expect changes in the ice mantle composition as desorption
due to protostellar heating occurs in the inner envelope.
How this affects the observed dust opacity is still unknown.
Unfortunately, NICMOS was not sensitive enough to detect
to detect background stars within $10$\as\ of the protostar.

	The underlying density and temperature distribution
affect the slope of the correlation through $P_n$.   
In order to analyze the effect of the temperature profile on 
the opacity ratio,
we first assume an isothermal approximation where 
every line-of-sight is assumed to have the same dust temperature
($T(r) =$ constant).
The calculated
opacity ratios for isothermal temperatures 
are shown as the solid curve in Figure 5.  For example,
the $850$ \micron\ ratio 
varies by a factor of $2.5$ for dust temperatures from $7$ to $14$ K. 
In reality, each line-of-sight through the protostellar 
envelope is non-isothermal (\S3).  We calculate $P_n(\theta)$ from
Equation 7 using the
density $(n(r))$ and dust temperature profiles $(T_d(r))$ determined
from the dust radiative transfer models for each line-of-sight ($\theta$). 
For each of the dust continuum models, we estimated the 
isothermal temperature
through the stellar lines-of-sight used
in the linear regression by solving Equation 7 for
the temperature and using the calculated line-of-sight
$P_n(\theta)$ from the radiative transfer model
\begin{equation}
T_{los}(\theta) = \frac{h\nu/k}{\ln\left(1 + \frac{2h\nu^3}{2.5\log_{10}(e) c^2 P_n(\theta)}\right)}
\;\; .
\end{equation}
This temperature corresponds to the single temperature that
characterizes a non-isothermal line-of-sight.
The average $\mean{T_{los}}$ is calculated by averaging $T_{los}$
for the 190 lines-of-sight used in the linear regression (\S4.1).
When the more realistic $T(r)$
from the radiative transfer models is included in the regression,
then the calculated opacity ratio is always below
the opacity ratio determined from isothermal lines-of-sight 
(Figure 5).  This is a systematic effect caused by a
monotonically decreasing temperature profile ($dT/dr < 0$; 
see Figure 2a) along each line-of-sight.  It is very important 
to account for temperature gradients when determining the opacity ratio
in the envelopes of Class 0 protostars.

	Figure 5 also graphically illustrates the
range of uncertainty introduced into the determination
of \kaprat\ due to different model opacity
assumptions and different physical models.  In
quoting our opacity ratio, we choose two limiting models
that characterize the range of \kaprat\ and which also
fit the observed $850$ \micron\ flux.  Those two
models are $2.8\times$Harvey BPL OH5 and
$2.8\times$Evans Shu OH8 in Table 1.
We find that 
$\frac{\kappa_{850}}{\kappa_{2.2}} =  (3.21 - 4.80)^{+0.44}_{-0.30} \times 10^{-4}$
and
$\frac{\kappa_{450}}{\kappa_{2.2}} =  (12.8 - 24.8)^{+2.4}_{-1.3} \times 10^{-4}$,
where the range in \kaprat\ corresponds to the value for each physical model.


   We may compare our value of the opacity ratio with previous
determinations toward dense cores (Figure 6).  Bianchi et al. (2003) observed
the starless core B68 at 850 and 1200 \micron\ with SCUBA and
SIMBA.  They employ a similar technique to compare the submillimeter
intensity and near-infrared colors.  However, they assume 
isothermality of the dust temperature and the extinction
law of Rieke \& Lebofsky (1986).  Since they determine
the opacity ratio with respect to the opacity at optical
wavelength, V, we must
use the Rieke \& Lebofsky $A_K/A_V = 1 / 8.9$ to convert from
$\kappa_{V}$ to
$\kappa_{2.2}$.  The resulting opacity ratios are
$\kappa_{850}/\kappa_{2.2} = 3.6 \pm 0.9 \times 10^{-4}$
and
$\kappa_{1200}/\kappa_{2.2} = 8.0 \pm 2.7 \times 10^{-5}$.
The $850$ \micron\ ratio is comparable to the ratio
we determined for B335.

    Similarly, Kramer et al. (2003) determined the
$850$ \micron\ opacity ratio toward 4 cores in the IC 5146
filament.  This study includes an analysis of
dust temperature variations between the cores in the filament.
Again we must use the Rieke \& Lebofsky
$A_K/A_V$ to convert to $\kappa_{2.2}$.  The four cores
have $\kappa_{850}/\kappa_{2.2}$ that range from
$1.9 \pm 0.2 \times 10^{-4}$ to $5.4 \pm 0.3 \times 10^{-4}$.
Kramer et al. also find evidence that the opacity ratio has an
inverse dependence on the dust temperature (see Figure 5).
The $\mean{\kappa_{850}/\kappa_{2.2}}$ from
all three studies is $3.7 \pm 1.4 \times 10^{-4}$. 
Unfortunately,
no opacity ratios have been determined previously in the literature
at 450 \micron\ toward low-mass dense cores. 
  

   Converting the observed opacity ratio to the submillimeter
opacity requires an estimate of the opacity at $2.2$ \micron .
A first approximation is
to average the $\kappa_{2.2}$ from many different theoretical
opacity models.  Averaging the $2.2$ \micron\ opacities
from the Ossenkopf \& Henning models (OH2, OH5, OH8; 1994), 
the opacities used in the multi-dimensional dust models
of Whitney et al. (private communication; e.g., Whitney et al. 2003),
the Mathis et al. opacity (1983), 
and the new theoretical 
opacities calculated by K. Pontoppidan (private communication) 
that match the Cores to Disk mid-infrared 
extinction law and ice features (Pontoppidan et al., in preparation), 
we find $\mean{\kappa_{2.2}} = 3800 \pm 700$ cm$^2$ g$^{-1}$ of dust.
Multiplying this number into the opacity ratios results
in the submillimeter opacities of
$\kappa_{850} = (1.18 - 1.77)^{+0.36}_{-0.24}$ and 
$\kappa_{450} = (4.72 - 9.13)^{+1.9}_{-0.98}$ cm$^2$ g$^{-1}$ of dust.
These opacities are plotted with theoretical curves in Figure 6.
We note that this crude average for $\kappa_{2.2}$ results in a large errorbar in
the calculated $850$ and $450$ \micron\ opacities because the $2.2$ \micron\
opacities vary by a factor of two among the different theoretical models.

The uncertainty in the $850$ and $450$ \micron\ opacity ratios and
opacities make our determinations
consistent with the empirical opacity law parameterized by Mathis (1990;
$\kappa = 13.16(\lambda / 250 \micron )^{-2}$; also parameterized 
in Kramer et al. 2003)
as well as the coagulated dust model with thin ice
mantles of Ossenkopf and Henning (OH5, 1994).  
The opacity model, OH8, that provides the best fit to the submillimeter
and intensity profile and SED is at the upper statistical
errorbar at 850 \micron\ and the lower bound of the models 
at 450 \micron .  Our results bracket the popular
theoretical opacities (OH5) at 850 \micron\ that 
have been used in dust continuum radiative
transfer modeling (e.g., Shirley et al. 2002,  Mueller et al. 2002,
Young et al. 2003, Shirley et al. 2005, Dunham et al. 2006, 
Doty et al. 2010).  

\subsection{Power-law index $\beta$}

   At far-infrared and submillimeter wavelengths $> 100$ \micron , 
the opacity
falls as a power-law with increasing wavelength 
($\kappa(\lambda ) \propto \lambda^{-\beta_{smm}}$).  Estimating $\beta_{smm}$
is a difficult problem.   The most traditional methods
have used modified blackbody fits to the SED or ratios of
submillmeter wavelengths to constrain $\beta_{smm}$ (e.g.
Visser et al. 1998, Shirley et al. 2000).  Both of 
these methods assume a single dust temperature which is not
an appropriate assumption for Class 0 protostars which have
strong temperature gradients (i.e., Fig 2).  Instead,
we use the derived opacity ratios at two submillimeter
wavelengths to constrain $\beta_{smm}$ between
450 and 850 \micron .  Since this method utilizes background
stars that are at least $15$\as\ from the central protostar,
the opacity ratios are probing the dust properties in the outer,
cold portions of the envelope.  This differs from the previous two
methods which use fluxes that include significant contributions
from warmer dust near the protostar.  
The opacity ratio $\beta_{smm}$ is given by
\begin{equation}
\beta_{smm} = \frac{ \ln \left( \frac{R_K \kappa_{450}/\kappa_{2.2}}
{R_K \kappa_{850}/\kappa_{2.2}} \right) }{\ln(850/450)} \; = 1.572 \,\ln \left( 
\frac{b_{450}}{b_{850}} \right) \;\;.
\end{equation}
Since this is a ratio, the exact value of $R_K$ cancels; 
however, the ratio is still sensitive to systematic
uncertainties such as flux calibration errors.  The opacity ratio
for B335 is $\beta_{smm} = (2.18 - 2.58)^{+0.30}_{-0.30}$.  

   If we compare our results to Bianchi et al. (2003)
which determined the opacity ratio at 850 and 1200 \micron , 
we find a severe discrepancy that illustrates the
importance of the calibration in determining $\beta$.   
The Bianchi $\beta_{smm} = 4.3 \pm 1.3$ is much higher
than our $\beta_{smm}$.  This anomalous result indicates
a systematic calibration problem at one or both wavelengths.
Great care must be taken when comparing observations made
with different instruments on different telescopes
through different observing conditions (e.g. SCUBA and SIMBA).
Observations taken with SCUBA simultaneously at $850$ and $450$ \micron\
avoid this problem since the observations
are taken in the same atmospheric conditions.  While a 
calibration error at 450 or 850 \micron\ could 
account for our $\beta_{smm} > 2$, we have taken
great care to assure a stable calibration between $850$ and
$450$ \micron\ by comparing the flux calibration of Uranus
taken on several nights surrounding the B335 observations.
The ratio of the $450$ to $850$ \micron\ 
flux conversion factors (see Jenness et al. 2002)
never vary by more than $10$\% during these time periods.

Observational evidence for submillimeter opacity indices
above two in the ISM exist.  
A detailed multi-wavelength study of the starless
core, TMC-1C, using
multiple methods to determine the opacity index find
that $1.7 \leq \beta \leq 2.7$ with a most likely
value near $\beta = 2.2$ (Schnee et al. 2010).
Another example is the PRONAOS 
(PROgramme NAtional d'Observations Submillim\'ertiques;
Lamarre et al. 1994)
balloon-borne experiment which finds that the opacity index
has an inverse temperature dependence with 
$\beta > 2$ for $T < 12.7$ K (Dupac et al. 2003).
The lower bound of our opacity index overlaps with
the PRONAOS opacity index curve ($\beta = 1/(0.40 + 0.0079T)$;
Dupac 2009) for the typical $\mean{T_{los}}$ found in 
our dust continuum models (see Figure 5).
The PRONAOS results are not unique 
as an inverse temperature dependence of $\beta$ and
opacity indices greater than two at low temperatures
has also been seen in far-infrared and submillimeter
observations from the ARCHEOPS balloon-borne
experiment (D\'esert et al. 2008).

The tendency of the opacity law toward $\beta = 2$
was noted from early submillimeter observations
and is thought to originate from behavior of the complex dielectric
function $\epsilon$ (Re($\epsilon$)=const, Im($\epsilon$)$\propto \nu$; Wickramasinghe 1967)
of the grains at wavelengths far from
resonances in the grain materials (e.g. Gezari et al. 1973).
None of the popular opacity models used in protostellar dust
continuum modeling or modeling of ISM dust
predict power-law indices greater than two
(i.e., OH2 $\beta_{smm} = 1.35$, OH5 $\beta_{smm} = 1.85$, 
OH8 $\beta_{smm} = 1.88$, WD5.5 $\beta_{smm} = 1.69$); 
see Shirley et al. 2005).  While our individual submillimeter
opacity constraints overlap the OH5 model at both $450$ and
$850$ \micron\ due to the uncertainty in the physical model
that best fits B335, the opacity index must be determined
using the same physical model at both wavelengths and 
the resulting $\beta_{smm}$ is too steep to be
consistent with the OH5 model.

There is a class of amorphous silicate dust
models which include phonon difference processes
(Disordered Charge Distributions and localized Two Level Systems; 
see Schl\"omann 1964 and Phillips 1987)
that result in $\beta_{submm} > 2$ at low temperatures 
(e.g., Meny et al. 2007).
These processes have been used to explain the anti-correlation
between $\beta_{smm}$ and temperature observed 
by the PRONAOS experiment (Boudet et al. 2005).
Our $\beta_{smm}$ range is consistent with the predicted opacity 
index from Boudet et al. (2005) and Meny et al. (2007)
for the typical $T_{los} < 10$ K derived from the dust models.
Ultimately, our results should be tested by reproducing this
analysis for observations with the new generation of submillimeter
cameras (e.g., SPIRE, LABOCA, SCUBA2) and using better constraints
from more sophisticated (e.g., multi-dimensional) dust continuum
models of B335.

\section{Conclusions}

We have determined
the opacity ratio from the slope of a plot of submillimeter intensity versus
near-infrared color excess
toward B335.  The submillimeter intensity along each 
line-of-sight in the correlation is corrected for
the non-isothermal temperature profile by the quantity $1/P_n$ which
is related to the density-weighted average Planck function.
We find opacity ratios of 
$\frac{\kappa_{850}}{\kappa_{2.2}} =  (3.21 - 4.80)^{+0.44}_{-0.30} \times 10^{-4}$
and
$\frac{\kappa_{450}}{\kappa_{2.2}} =  (12.8 - 24.8)^{+2.4}_{-1.3} \times 10^{-4}$
for a ratio of total to selective K-band extinction of $R_K = 1.59 \pm 0.12$.
The range in values
corresponds to the uncertainty in the physical model for the envelope
of B335.
The submillimeter opacity power-law index is
$\beta_{smm} = (2.18 - 2.58)^{+0.30}_{-0.30}$.    
For an average $2.2$ \micron\ opacity of $3800 \pm 700$ cm$^2$g$^{-1}$,
we find an opacity at 850 and 450 \micron\ of 
$\kappa_{850} = (1.18 - 1.77)^{+0.36}_{-0.24}$ and 
$\kappa_{450} = (4.72 - 9.13)^{+1.9}_{-0.98}$ cm$^2$g$^{-1}$.
These opacities statistically
agree with the popular theoretical ratios of Ossenkopf and Henning for
coagulated ice grains with thin mantles $(0.65 - 0.97) \kappa^{\rm{OH}5}_{850}$ 
at 850 \micron ; however, our derived opacity index ($\beta_{smm}$)
is steeper than predicted by the OH5 model ($\beta_{OH5} = 1.85$).
This comparison of near-infrared color excess and submillimeter emission probes
the opacity on scales of $15$\as\ to $75$\as , and does not find
evidence for a large scale variation in the opacity on those scales.  
We confirm a disagreement between the best fitted dust radiative
transfer model and the best-fitted molecular line radiative
transfer model.  Improvements in the estimate of the opacity ratios
and submillimeter opacities may be made with more sophisticated,
milti-dimensional modeling
of the dust continuum emission such as variable dust opacities 
in the inner envelope 
where desorption of CO and other molecules may change the optical
constants of grains.  The techniques used in this analysis
should be applicable to far-infrared and submillimeter observations
of B335 with the \textit{Herschel Space Observatory}.
With the commissioning of new, sensitive bolometer cameras,
such as LABOCA and SCUBA-2, combined with observations with large format 
infrared CCDs on large aperture telescopes (JWST, Keck, etc.),
it will be possible to extend this method to study the dust opacity 
ratio around other Class 0 protostars.

\section*{Acknowledgments}

We sincerely thank the referee for a very thorough and careful
reading of the manuscript.
We thank Brandon Kelly for comments that improved this paper.
Guest User, Canadian Astronomy Data Centre, which is operated by the Dominion 
Astrophysical Observatory for the National Research Council of Canada's 
Herzberg Institute of Astrophysics. 
KMP is supported by NASA through Hubble  
Fellowship grant \#01201.01 awarded by the Space
Telescope Science Institute, which is operated by the Association of  
Universities for Research in Astronomy, Inc., for NASA, under contract
NAS 5-26555. 
Additional support came from NASA Origins grant NNG04GG24G  and the
\textit{Spitzer} Legacy Science Program, provided by NASA through contracts
1224608 and 1230779 issued by the Jet Propulsion Laboratory,
California Institute of Technology, under NASA contract 1407.
to NJE.




\begin{figure}
\figurenum{1}
\epsscale{1.1}
\plotone{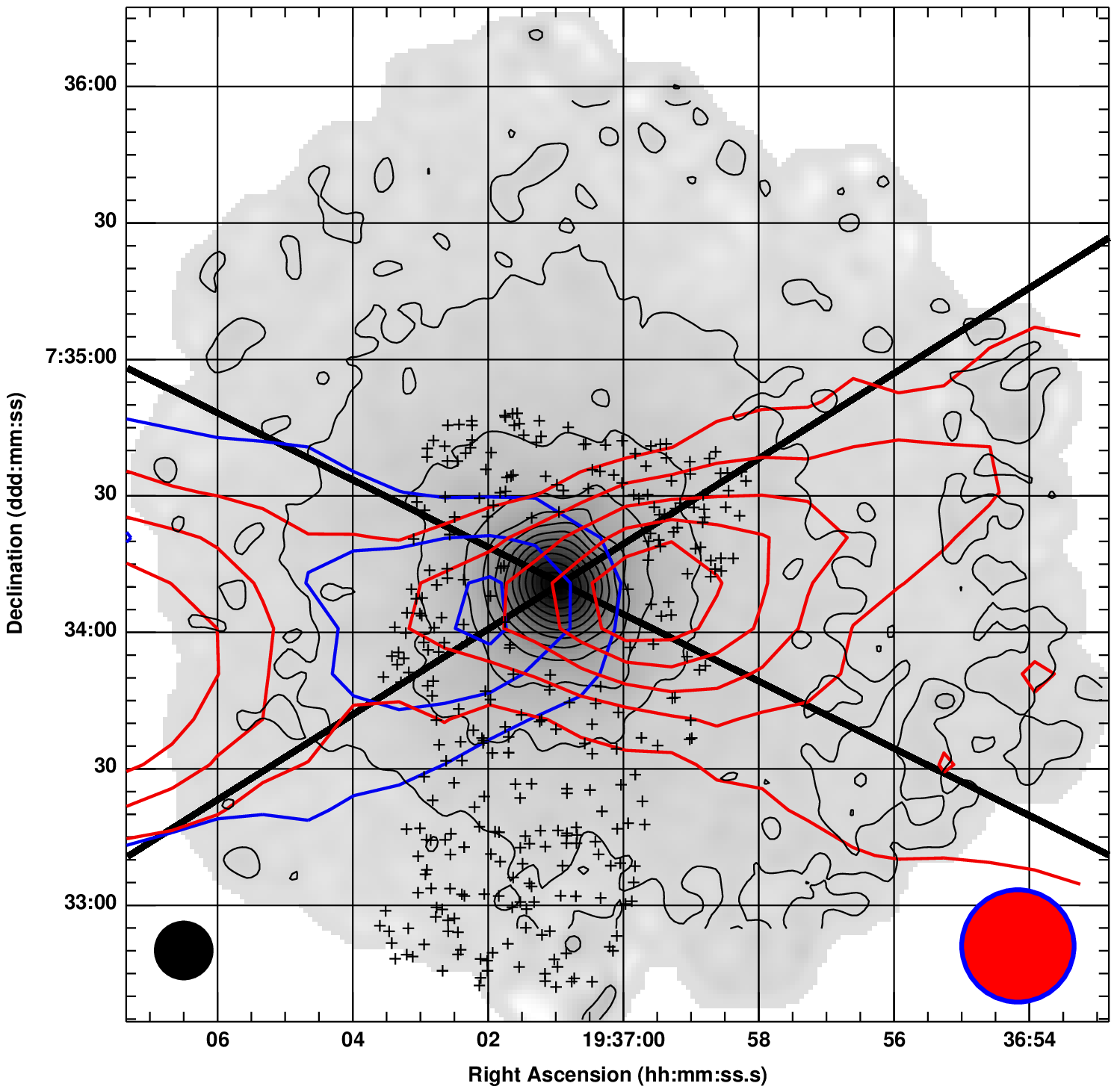}
\figcaption{Greyscale 850 \micron\ image of B335 with positions
of background stars observed with NICMOS indicated by the
small black crosses.  All positions are in the epoch J2000.0.
The $850$ \micron\ (black) contours starts at $2 \sigma$
(30 mJy/beam) and then are spaced at $10$\% of the peak ($101$ mJy/beam).
The red and blue contours trace the outflow
wings derived from CO J $= 2 \rightarrow 1$ observations and
start at $1.5$ K km/s and increase by $1.5$ K km/s (Stutz et al. 2008).  
The extent of the outflow
cavity used to exclude background stars is shown by the two solid black lines.
The SCUBA beam (lower left) and SMT CO beam (lower right) are displayed.}
\end{figure}


\begin{figure}
\figurenum{2}
\epsscale{1.0}
\plotone{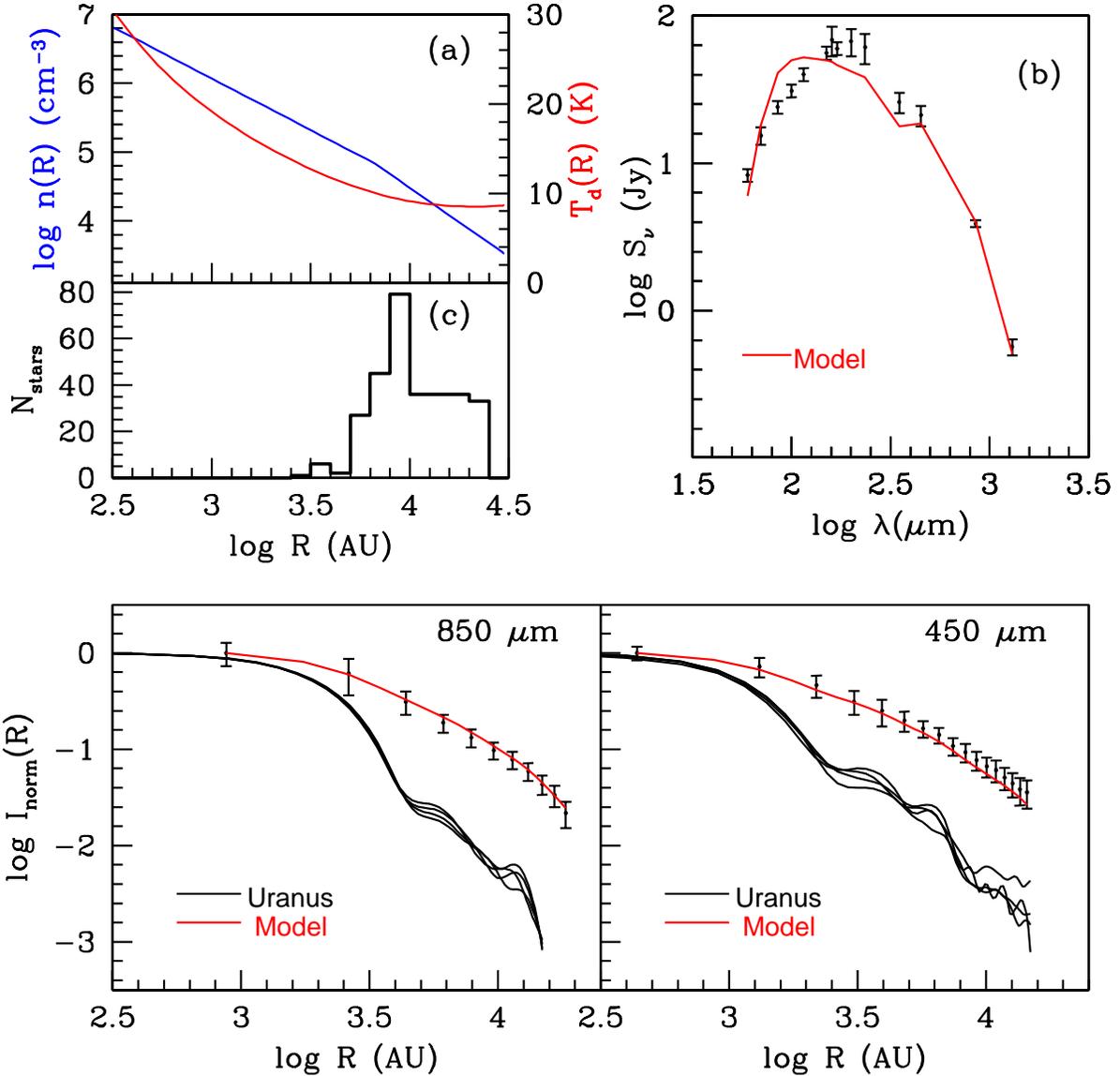}
\figcaption{Updated best-fit dust model for B335,
$2.4\times$Harvey BPL OH8.  Panel (a)
displays the scaled-Harvey broken power-law density profile
(blue) and the resulting dust temperature profile (red).
The fit to the SED is shown in panel (b).  The histogram
showing the location of background stars that are used to
constrain the opacity ratio is shown in panel (c).  The
distribution is strongly peaked just below $10^4$ AU.
The fit to the submillimeter intensity profiles at
850 and 450 \micron\ are shown in the bottom panels.
The red curves are the dust model profiles while the black
curves are the beam profiles determined from Uranus
observations bracketing the B335 observations.  The intensity
errorbars account for statistical uncertainty in the
intensity as well as azimuthal variations in intensity within
each annulus.}
\end{figure}


\begin{figure}
\figurenum{3}
\epsscale{0.9}
\plotone{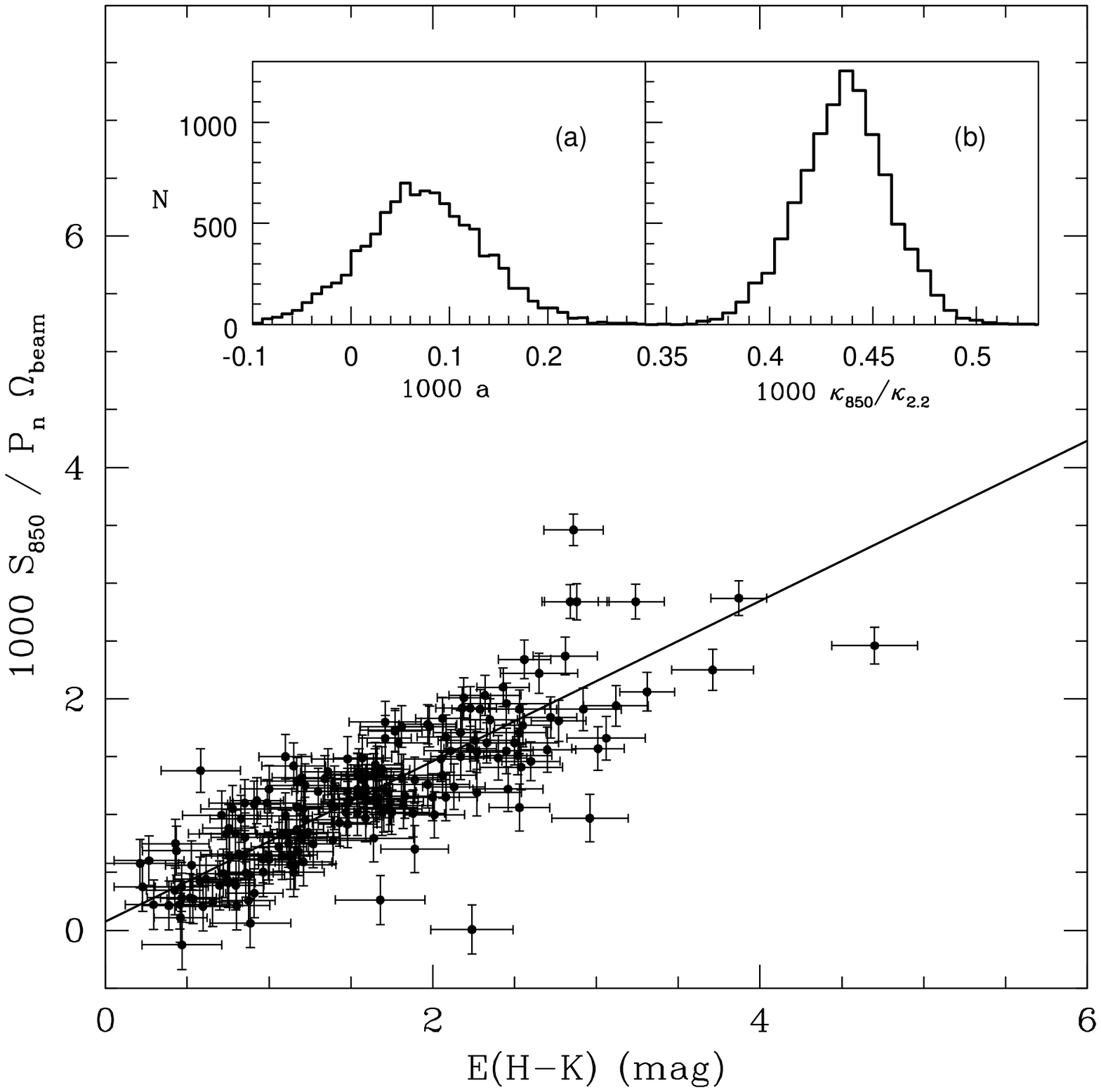}
\figcaption{The 850 \micron\ intensity plotted versus (H - K)
color excess.  $P_n$ is related the density-weighted Planck function.
The solid line is the linear regression with the mean slope
and intercept from the Posterior distributions.
Histograms of the intercept (a) and slope (b) distributions are shown in
the insets.  $\kappa_{850}/\kappa_{2.2} = b_{850} / R_K$ where we
have assumed $R_K = 1.59$ (see \S2).}
\end{figure}


\begin{figure}
\figurenum{4}
\epsscale{0.9}
\plotone{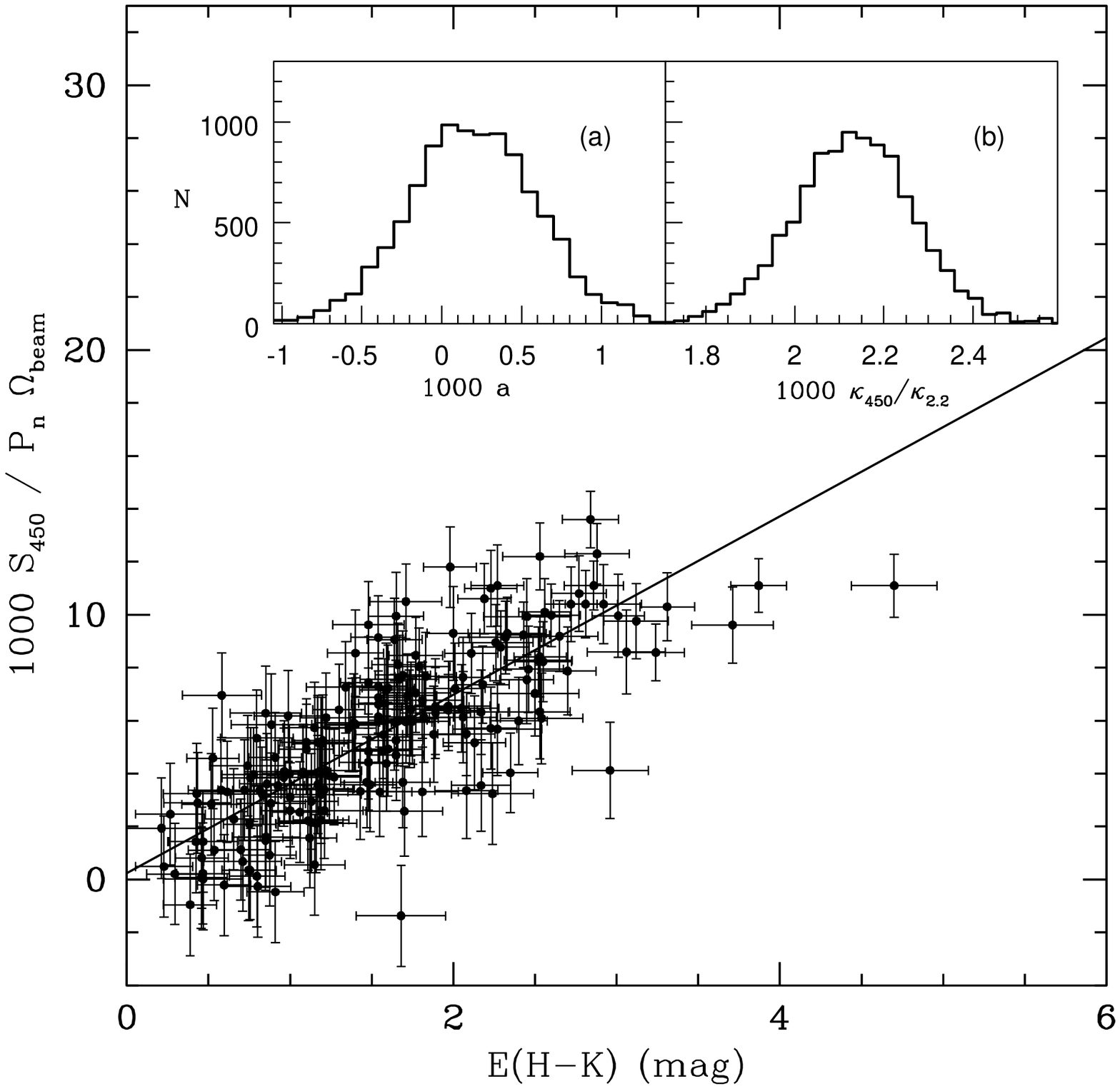}
\figcaption{The 450 \micron\ intensity plotted versus (H - K)
color excess.  $P_n$ is related to the density-weighted Planck function.
The solid line is the linear regression with the mean slope
and intercept from the Posterior distributions.
Histograms of the intercept (a) and slope (b) are shown in
the insets.  $\kappa_{450}/\kappa_{2.2} = b_{450} / R_K$ where
we have assumed $R_K = 1.59$ (see \S2).}
\end{figure}


\begin{figure}
\figurenum{5}
\epsscale{0.9}
\plotone{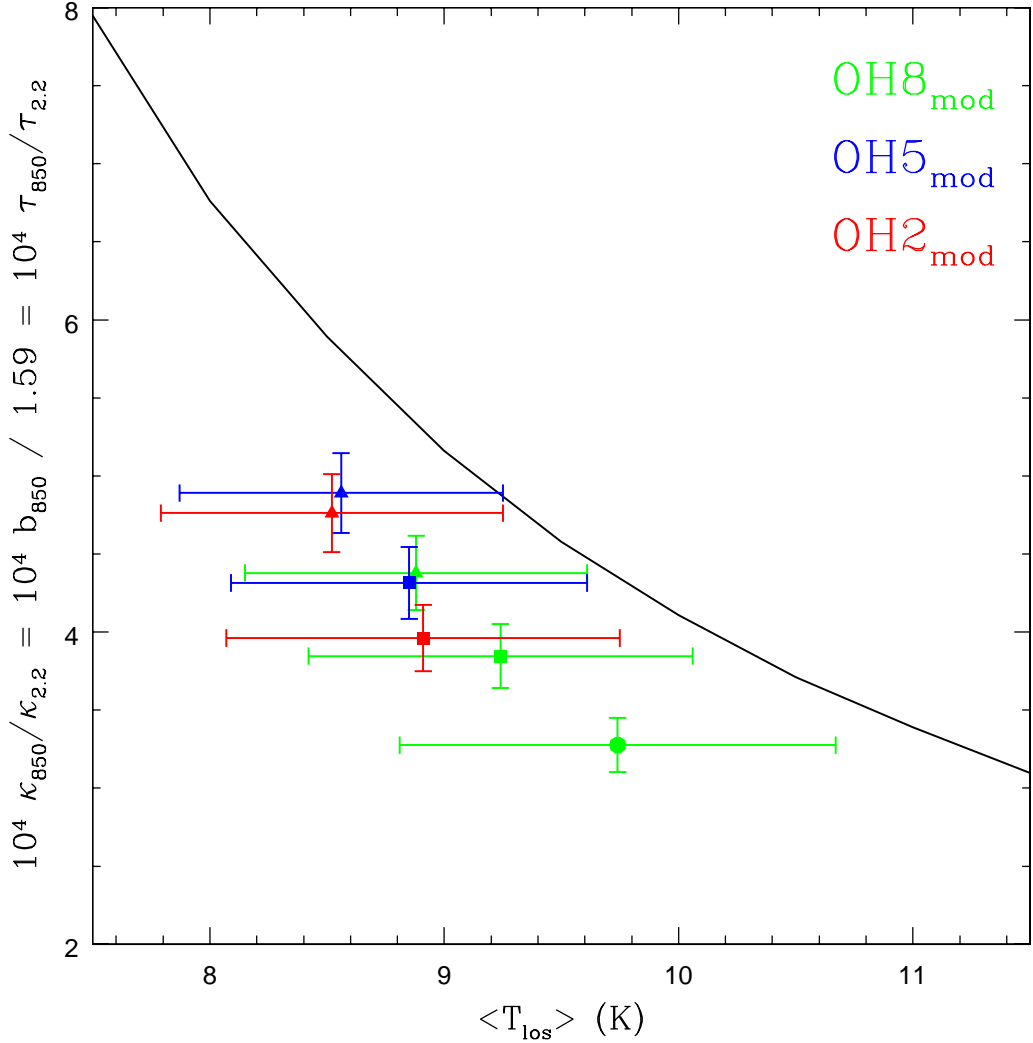}
\figcaption{$\kappa_{850}/\kappa_{2.2}$ versus the average line-of-sight
dust temperature.  $\kappa_{850}/\kappa_{2.2}$ is determined from the slope
of linear regressions ($b_{850}$ 
see Figures 3 and 4) divided by $R_K = 1.59$ (see \S2)
for different physical models ($n(r)$, $T(r)$).
The solid line is for dust models with an isothermal
envelope ($T(r) =$ constant).  The plotted points are for various
dust models with calculated $T(r)$.  Symbols refer to: scaled Harvey 
broken-power law (triangles), scaled Shu model (squares),
and Evans-Shu model (circle).  Color indicates the dust opacity
model used to calculate $T(r)$: OH8$_{mod}$ (green), OH5$_{mod}$ (blue), 
and OH2$_{mod}$ (red) (\S 3). 
All models except for the Evans-Shu
model are scaled in density to match the 
observed $850$ \micron\ flux.}
\end{figure}


\begin{figure}
\figurenum{6}
\epsscale{0.9}
\plotone{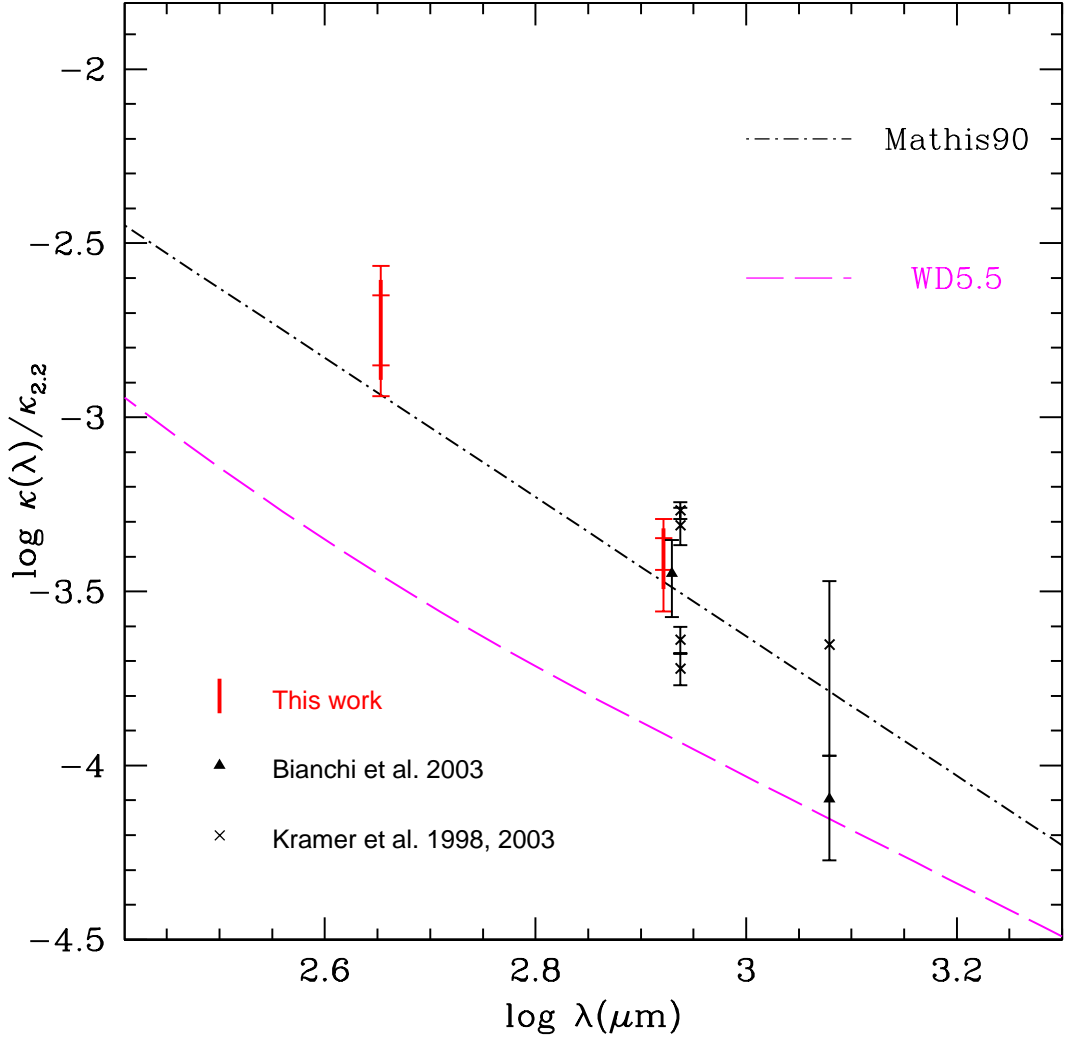}
\figcaption{Theoretical kappa ratios with the observed opacity
ratios from this work (B335, shown in red), Bianchi et al. (B68, 2003), and 
Kramer et al.(IC 5146, 1.2mm 1998, 850 \micron\ 2003).
The Kramer et al. 850 \micron\ points have been shifted
slightly in wavelength for clarity.  
WD $=$ Weingartner \& Draine (2001)
for $R_V = 5.5$.  
Mathis refers to the parametrization by 
Mathis (1990) of the ISM empirical dust model.  OH model profiles
are not included in this Figure since self-consistent 
scattering opacities were not
determined by Ossenkopf \& Henning (1994).}
\end{figure}


\begin{figure}
\figurenum{7}
\epsscale{0.9}
\plotone{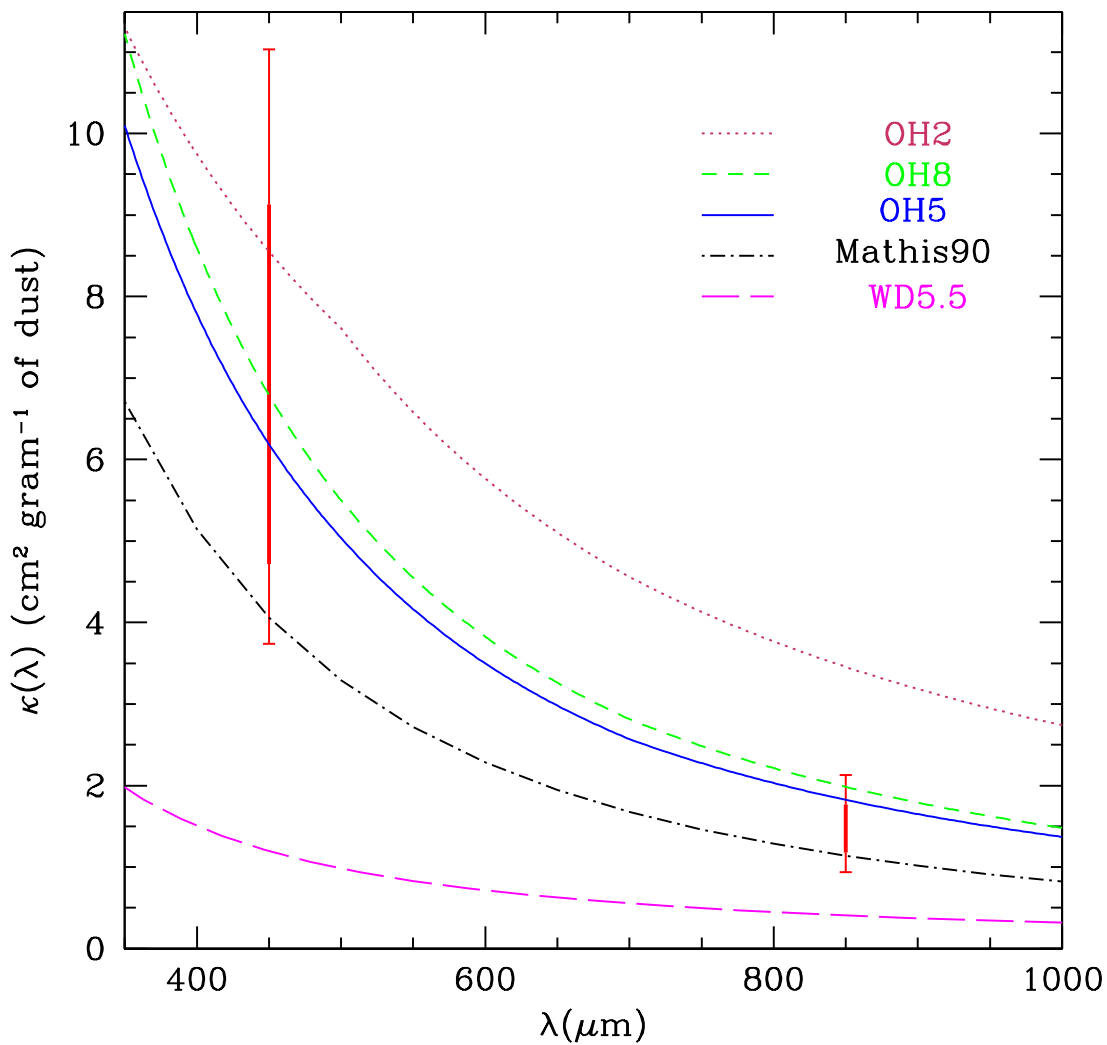}
\figcaption{Our constraints on the submillimeter dust opacity
assuming and absolute value of the opacity at $2.2$\micron\
of $3800 \pm 700$ cm$^2$ g$^{-1}$.  WD $=$ Weingartner \& Draine (2001)
for $R_V = 5.5$.  OH $=$ Ossenkopf \& Henning (1994)
for (2) no ice mantles, (5) thin ice mantles, and (8) thick
ince mantles (N. B., the Young \& Evans 2005 modification to the OH opacities
does not affect the submillimeter opacities).  Mathis referes to the parametrization by 
Mathis (1990) of the ISM empirical dust model.}
\end{figure}





\begin{deluxetable}{lccccccccc}
\rotate
\footnotesize
\tablecolumns{10}
\tablecaption{Properties of Selected Radiative Transfer Models \label{tab1}}
\tablewidth{0pt}
\tablehead{
\colhead{Physical Model\tablenotemark{a}} &
\colhead{$\chi^2_{I_{450}}$} &
\colhead{$\chi^2_{I_{850}}$} &
\colhead{$\chi^2_{SED}$} &
\colhead{$L_{>60}^{mod}$ (\lsun )\tablenotemark{b}} &
\colhead{$\mean{T_{los}}$ (K)} &
\colhead{$10^4b_{850}$} &
\colhead{$10^4a_{850}$} &
\colhead{$10^4b_{450}$} &
\colhead{$10^4a_{450}$}
}
\startdata
$2.4\times$Harvey BPL OH8\tablenotemark{c}	& 0.36	& 0.68	& 29.55	& 3.0	& 8.88 $\pm$ 0.73	&
	6.93 $\pm$ 0.37 & 0.74 $\pm$ 0.65 & 33.7 $\pm$ 2.1 & 2.3 $\pm$ 3.7 \\
$2.8\times$Harvey BPL OH5			& 0.23	& 1.19	& 32.80	& 2.6	& 8.56 $\pm$ 0.69	&				7.63 $\pm$ 0.40 & 0.67 $\pm$ 0.69 & 39.4 $\pm$ 2.3 & 1.6 $\pm$ 4.2 \\
$1.4\times$Harvey BPL OH2			& 0.26	& 2.01	& 43.20	& 1.8	& 8.52 $\pm$ 0.73	&				7.34 $\pm$ 0.39 & 1.18 $\pm$ 0.68 & 36.5 $\pm$ 2.3 & 6.4 $\pm$ 4.3 \\
$2.8\times$Evans SHU OH8\tablenotemark{d}	& 4.06 	& 12.37	& 32.29	& 2.7	& 9.24 $\pm$ 0.82	&				6.00 $\pm$ 0.32 & 1.11 $\pm$ 0.56 & 26.2 $\pm$ 1.7 & 5.8 $\pm$ 3.2 \\
$11.0\times$Harvey BPL WD5.5			& 0.76	& 1.54	& 17.91	& 2.7	& 9.18 $\pm$ 0.71	&				7.61 $\pm$ 0.37 & $-$0.75 $\pm$ 0.65 & 39.2 $\pm$ 2.1 & $-$1.0 $\pm$ 0.4 \\
$3.1\times$Evans SHU OH5			& 1.98	& 10.42	& 44.57	& 2.1	& 8.85 $\pm$ 0.76	& 				6.73 $\pm$ 0.36 & 1.06 $\pm$ 0.64 & 32.0 $\pm$ 2.0 & 5.1 $\pm$ 3.8 \\
$1.6\times$Evans SHU OH2			& 0.75	& 6.00	& 43.41	& 1.5	& 8.91 $\pm$ 0.84	&				6.18 $\pm$ 0.33 & 1.67 $\pm$ 0.60 & 27.1 $\pm$ 1.8 & 10.8 $\pm$ 3.5 \\
$14.0\times$Evans SHU WD5.5			& 0.23	& 17.91 & 37.45 & 2.3	& 9.35 $\pm$ 0.72	&				6.85 $\pm$ 0.36 & $-$0.16 $\pm$ 0.62 & 33.4 $\pm$ 1.9 & $-$5.2 $\pm$ 3.3 \\
$1.0\times$Evans SHU OH5			& 7.72	& 15.12	& 53.49	& 1.9	& 9.74 $\pm$ 0.93	&				5.11 $\pm$ 0.27 & 1.33 $\pm$ 0.47 & 20.4 $\pm$ 1.4 & 7.0 $\pm$ 2.7 \\
\enddata
\tablenotetext{a}{Harvey BPL = Harvey et al. (2003b) broken power-law $n(r)$ and Evans SHU = 
Evans et al. (2005) Shu-infall $n(r)$.  The numbers refer to scaling factors
multiplied into the density or the opacity.  All models except $1.0\times$Evans SHU OH5
are scaled to match the observed flux at $850$ \micron .}
\tablenotetext{b}{Luminosity integrated from $\lambda \geq 60$ \micron .  The B335 
SED is published in Shirley et al. (2002) plus additional points from \textit{Spitzer
Space Telescope} observations at $70$ \micron\ 
($S = 15.4 \pm 2.1$ Jy in a $70$\as\ aperture)
and $160$ \micron\ ($S = 68.7 \pm 15.6$ Jy in a $100$\as\ aperture) from Stutz et al. (2008).}
\tablenotetext{c}{Best-fitted dust continuum model.}
\tablenotetext{d}{Best-fitted molecular model that matches flux at $850$ \micron .}
\tablenotetext{e}{$a$ and $b$ refer the the intercept and slope of the linear regression at $850$ and $450$ \micron .}
\end{deluxetable}

\end{document}